\documentclass[aip,preprint,a4paper,jap]{revtex4-1}
\usepackage{amsmath,graphicx}
\newcommand{\un}[1]{\ensuremath{\, \mathrm{#1}}}

\begin{document}
\title{Interplay of bulk and surface properties for
steady-state measurements of minority carrier lifetimes}
\date{\today}
\author{Marko Turek}
\email{marko.turek@csp.fraunhofer.de}
\affiliation{Fraunhofer-Center for Silicon Photovoltaics CSP}
\begin{abstract}
The measurement of the minority carrier lifetime is a
powerful tool in the field of semiconductor
material characterization as it is very sensitive to
electrically active defects. Furthermore, it is applicable
to a wide range of samples such
as ingots or wafers. In this work, a systematic
theoretical analysis of the steady-state approach 
is presented.
It is shown how the measured lifetime relates to
the intrinsic bulk lifetime for a given material quality, sample
thickness, and surface passivation. This
makes the bulk properties experimentally accessible by
separating them from the surface effects.
In particular, closed analytical solutions
of the most important cases, such as
passivated and unpassivated wafers and blocks
are given.  Based on these results,
a criterion for a critical sample thickness is given
beyond which a lifetime measurement allows deducing the
bulk properties for a given surface recombination.
These results are of particular
interest for semiconductor material diagnostics especially for
photovoltaic applications but not limited to this field.

\end{abstract}
\maketitle

\section{\label{sec:Introduction}Introduction}
Minority carrier lifetime measurements have been proven
to be an essential material characterization technique
for semiconductors. The determination of the
lifetime provides information on electrically
active defects in a very sensitive way. It is possible
to detect defect concentrations down to 
$10^{10}{\rm{cm^{-3}}}$ with contactless methods
at room temperature.\cite{Schroder06,Rein05}

There are different approaches which are based
on either a dynamic process taking place after a pulsed carrier
generation (transient measurement)
or a steady-state measurement during a continuous carrier generation.
In either case, the measured lifetime $\tau_{\rm eff}$
incorporates both bulk and surface recombination processes.
However, a typical requirement on the lifetime measurement is that either
the material properties, i.e. bulk lifetime $\tau_{\rm b}$, or
the surface properties, i.e. surface recombination velocity $S$,
can be unambiguously deduced. The first information
on the bulk allows, for example, an optimization
of the production process of the semiconductor, i.e. the
crystallization. On the other hand, the second information
on the surface recombination plays an important role, for example,
for a device optimization like the development of the passivation
layers for solar cells.
Therefore, it is inevitable to know the precise relationship
between the measured lifetime $\tau_{\rm eff}$,
the bulk lifetime $\tau_{\rm b}$ and the surface recombination
velocity $S$, i.e. the
function $\tau_{\rm eff} = \tau_{\rm eff}(\tau_{\rm b},S)$.
Typically, this relationship depends
on sample properties such as its thickness $d$
as well as the spectrum
of the light generating the carriers and the corresponding
absorption length $\alpha^{-1}$.

For transient measurements and
equal surface recombination velocities on the front and back side of the sample ($S_0=S_1=S$),
the most frequently used functional relation is given
by $\tau_{\rm eff}(\tau_{\rm b},S,d) = (\tau_{\rm b}^{-1} + \tau_{\rm s,tr}^{-1})^{-1}$.
Here, the so-called surface recombination lifetime $\tau_{\rm s,tr}$ is
given by the phenomenological approximation
$\tau_{\rm s,tr} \approx d/(2 S) + d^2/(\pi^2 D)$
while $D$ stands for the diffusion constant of the minority carriers.
This approximation is a superposition of two asymptotic
solutions of the partial differential equation for the
excess carrier density $p$
corresponding to a transient measurement.
The first term is exact in the limit of
very low surface recombination and thin samples while
the second term is found for very high surface
recombination and thick samples.\cite{Luke87,Grivickas89}
To obtain this result, only the fastest decaying mode
of the density after a short illumination pulse
is considered which leads to a transcendental
equation for $\tau_{\rm s,tr}$ that can be solved in the $Sd/D \ll 1$
and $S d / D \gg 1$ limits, respectively. In subsequent works, this
approach has been extended to the case of unequal
$S_0 \neq S_1$ and a systematic approach based on
the relevant dimensionless parameters has been
presented.\cite{Kousik92,Otaredian93,Sproul94}

The second widely-used approach relies on a steady-state
measurement. A mathematical model for this case
can be formulated by an ordinary differential equation, i.e.
a time independent diffusion equation for the excess
carrier density $p$. Despite a number of assumptions
that have to be made in order to validate this model,
the solution to this
equation can give rather accurate results
for the carrier density.\cite{DeVore56,Duggan81}
However, it has been pointed out that the transient
approach and the steady-state approach can lead
to different results for $\tau_{\rm eff}$ in
certain cases.\cite{Nagel99b,Brody03} Furthermore, a number
of experimental techniques
are based on a steady-state of the sample
created by a long illumination, such as lifetime
measurements using photoluminescence (PL) or the
determination of the 
(quasi) steady-state photoconductance (QSSPC).\cite{Sinton96a}
In addition, the dependence of $\tau_{\rm eff}$
on the absorption length $\alpha^{-1}$ of the light
used for the carrier generation as well as the influence
of the surface recombination for intermediate values of $S$ is
of interest. 

A common approach to determine $\tau_{\rm b}$ experimentally
is to prepare
the samples such that the surface properties are negligible.
This can be achieved by a surface passivation which 
significantly reduces the surface recombination velocities.\cite{Aberle00}
In this case, the bulk lifetime is measured directly,
i.e. $\tau_{\rm eff} \approx \tau_{\rm b}$. For small surface
recombination velocities, a correction based on the
limiting expression $\tau_{\rm s} \approx d/(2S)$ can be employed
to improve the accuracy of $\tau_{\rm b}$.\cite{Schmidt97}
On the other hand, a method for determining $\tau_{\rm b}$
on thick unpassivated samples, i.e. blocks, has been recently
proposed.\cite{Bowden07} Furthermore, the feasibility of measuring
bulk properties on thin unpassivated wafers has been
investigated, both experimentally and numerically.\cite{Sinton04b,Bothe10}

In this work, we will present a systematic analysis
of the relation between the measured lifetime $\tau_{\rm eff}$,
the bulk property $\tau_{\rm b}$ and the surface recombination
velocity $S$ that is valid for the
steady-state approach. Our study is based on the function
$\tau_{\rm eff} = \tau_{\rm eff}(\tau_{\rm b},d,S_0,S_1,\alpha)$
that can be derived from the solution of the
time independent diffusion equation, see Sec.~\ref{sec:diffusion}.
It generalizes special cases obtained
earlier\cite{Sinton04b,Bowden07,Bothe10}
for the steady-state regime and furthermore allows
investigating all practically relevant sample types
within the same framework. In particular,
our study extends the previously proposed results for thin
wafers on one side and thick samples such as blocks on the
other side to samples of arbitrary thickness.
The qualitative behaviour of this
solution with respect to a variation in the parameters
$\tau_{\rm b}$, $d$, $S_{0,1}$, and $\alpha$ will be discussed in
Sec.~\ref{sec:application}.
Approximations valid for the relevant
cases of homogeneous generation (i.e. $\alpha \to 0$),
unpassivated samples (i.e. $S \to \infty$), and large
sample thickness (i.e. $d \to \infty$) will also be
given in Sec.~\ref{sec:application}.
These results allow for a straightforward interpretation
of the data obtained in lifetime measurements.
In Sec.~\ref{sec:asymptotic} a systematic approach is
presented that allows extracting all relevant asymptotic limits
with respect to the sample parameters $\tau_{\rm b}$,
$S$, $d$, and $\alpha$. As a direct application
of these asymptotic expressions a
criterion for the sample thickness beyond which
a reliable determination of the bulk lifetime is
possible will be derived.
It will be shown how this critical thickness $d_{\rm crit}$
depends on $S$, $\tau_{\rm b}$ and $\alpha$. As a second application
the correct expression for
$\tau_{\rm eff}$ in the limit of increasingly better
material quality, i.e. $\tau_{\rm b} \to \infty$, will be deduced 
and compared
to a phenomenological result published ealier.\cite{Sinton04b,Bothe10}
Finally, we will discuss the experimental conditions which allow
the separation of bulk and surface properties in an unambiguous
manner.

\section{\label{sec:diffusion}Diffusion model}
The measurable carrier lifetime $\tau_{\rm eff}$ for a steady-state
application is given by
\begin{equation}
\label{eq:definition_tau_eff}
\tau_{\rm eff}=\frac{N_p}{G}
\equiv \left( \int\limits_{-d/2}^{d/2} {\rm d}x \, p(x) \right)
\left/ \left( \int\limits_{-d/2}^{d/2} {\rm d}x \, g(x) \right) \right.
\end{equation}
with $N_p$ being
the total number of minority excess
carriers in the sample and $G$ the total
generation rate in the bulk, respectively.
For the steady state approach the generation rate per volume
is time independent and given by
\begin{equation}
g(x) = \frac{P \lambda}{A h c} (1-r) \frac{\alpha {\rm e}^{-\alpha d/2}}{1-r^2 {\rm e}^{-2 \alpha d}}
\left( 1 + r {\rm e}^{\alpha (2 x-d)} \right) {\rm e}^{-\alpha x}
\end{equation}
with $P$ being the power of the illumination, $A$ the area of the sample,
$\lambda$ the wavelength, $\alpha^{-1}$ the absorption length,
and $r$ the reflection coefficient.\cite{Luke87}
The diffusion equation and the boundary conditions, that describe the
carrier density $p(x)$, read
\begin{equation}
\label{eq:ODE}
\frac{{\rm d}^2 p}{{\rm d} x^2} - \frac{p(x)}{D \tau_{\rm b}}
+ \frac{g(x)}{D} = 0
\, , \;
\left. \frac{{\rm d} p(x)}{{\rm d} x} \right|_{\pm \frac{d}{2}} = 
\mp \frac{S_{1,0}}{D} \, p\left(\pm \frac{d}{2}\right) .
\end{equation}

Integrating the diffusion equation
straightforwardly yields $G = R + A [S_0 p(-d/2) + S_1 p(+d/2)]$
with $R$ being the total recombination rate in the bulk.
This allows for a simple interpretation of the
surface recombination velocities: all
generated carriers that are not recombining in the bulk
have to recombine at the surface and these processes are
characterized by $S_{0,1}$ and $p(\pm d/2)$. The solution to Eq.~(\ref{eq:ODE})
is given by
\begin{equation}
p(x)= c_1 {\rm e}^{x/l} + c_2 {\rm e}^{-x/l} + 
c_3 {\rm e}^{\alpha x} + c_4 {\rm e}^{-\alpha x}
\end{equation}
with $l\equiv \sqrt{\tau_{\rm b} D}$ being the diffusion
length.\cite{DeVore56,Duggan81,Brody02,Brody03,Schroder06}
The prefactors $c_{1-4}$ are obtained by inserting this solution into
the diffusion equation and the corresponding boundary conditions,
Eq.~(\ref{eq:ODE}).
Integration of the carrier density $p(x)$ and the generation rate
$g(x)$ according
to Eq.~(\ref{eq:definition_tau_eff}) then yields the function
$\tau_{\rm eff} = \tau_{\rm eff}(\tau_{\rm b}, S, d, \alpha)$.

The major assumptions for this model are:
\begin{enumerate}
\item lateral
 homogeneity of the sample allowing for a one-dimensional
 description;
\item constant, i.e. injection independent,
 material properties $\tau_{\rm b}$, $S$, and diffusion constant $D$;
\item monochromatic illumination with a single
 wavelength $\lambda$ and corresponding absorption depth $\alpha^{-1}$;
\item large penetration depth of the sensor measuring the total number
 of carriers $N_p$ in comparison to absorption length $\alpha^{-1}$ and
 diffusion length $l$ .
\end{enumerate}
The lateral homogeneity of the material has to be given on length scales
larger than the diffusion length.
With this condition fullfilled and
given that the measurement is performed in a steady state, the
variations in the material parameters are due to the
position dependent carrier density $p(x)$ only.
Furthermore, the model could easily be extended beyond the monochromatic case
by linear superposition. However, this would make the analysis technically
more complex while for the purpose of this work there are
no significant additional insights to be gained.

In order to get a deeper insight into the relationship
between the lifetimes, the surface properties and the sample
geometry we state the result for the most relevant cases.
First, in case of equal surface recombination, i.e.
$S_0=S_1=S$, one finds
\begin{equation}
\label{eq:tau_eff_equal_S}
\tau_{\rm eff}^{S_0=S_1} = \frac{\tau_{\rm b}}{1-\alpha^2 l^2} \left[ 1 -
 \alpha l \frac{\alpha l + \frac{Sl}{D} \coth \frac{\alpha d}{2}}
 {1+\frac{S l}{D} \coth \frac{d}{2 l}} \right] 
\end{equation}
This solution is valid for any sample thickness $d$, absorption length
$\alpha^{-1}$ and surface recombination velocity $S$.
Second, the result for homogenous generation
but with non-equal surface recombination velocities is presented
\begin{widetext}
\begin{equation}
\label{eq:tau_eff_zero_alpha}
\tau^{\alpha \to 0}_{\rm eff} = 
\tau_{\rm b} \left[ 1- \frac{\frac{(S_0+S_1)d}{D} \frac{l^2}{d^2} \sinh\left[\frac{d}{l}\right]
 + 4 \frac{S_0 d}{D} \frac{S_1 d}{D} \frac{l^3}{d^3} \sinh^2\left[\frac{d}{2l}\right]}
 {\frac{(S_0+S_1)d}{D}\frac{l}{d} \cosh\left[\frac{d}{l}\right]
  + \left(1+\frac{S_0 d}{D} \frac{S_1 d}{D} \frac{l^2}{d^2}\right)
  \sinh\left[\frac{d}{l}\right]} \right] \, .
\end{equation}
\end{widetext}
This result applies to wafers with a thickness being small enough such that
a homogeneous generation $g(x)=g$ is given, i.e. $\alpha d \ll 1$.

It is worthwhile to note, that one could in principle
determine a surface recombination time $\tau_{\rm s}$ from
Eq.~(\ref{eq:tau_eff_equal_S})
by applying $\tau_{\rm s}^{-1} = \tau_{\rm eff}^{-1} - \tau_{\rm b}^{-1}$.
However, the $\tau_{\rm s}$ obtained in this
way is in general 
not independent of $\tau_{\rm b}$ (or, similarly, of $l=\sqrt{D \tau_{\rm b}}$).
This is different from the case of
the approximate solution obtained in the transient approach.

In a typical experiment, the quantities $\tau_{\rm eff}$,
$\alpha$, and $d$ are known or measurable while $\tau_{\rm b}$ and $S$
are unknown. This implies that the stated relationship
$\tau_{\rm eff} = \tau_{\rm eff}(\tau_{\rm b}, d, S, \alpha)$,
i.e. Eq.~(\ref{eq:tau_eff_equal_S}),
in general allows for the determination of the two unknown
quantities if two independent measurements are performed
with either $d$ or $\alpha$ being varied.
For example, a well known approach to determine $S$ is to
set up an experiment where wafers of the same material and
surface treatment but with different thickness are measured.\cite{Rein05}
However, as will be shown in the next section, a careful
choice of range for these thicknesses has to be made if
both $S$ and $\tau_{\rm b}$ are to be determined within the same
experiment.

\section{\label{sec:application}Application to wafers and blocks}
For the following discussion of the resulting relation
$\tau_{\rm eff} = \tau_{\rm eff}(\tau_{\rm b}, S, d)$
we focus on the case where $|S_0 - S_1|/(S_0+S_1) \ll 1$ which
is reflected by Eq.~(\ref{eq:tau_eff_equal_S}).
The rather general result (\ref{eq:tau_eff_equal_S})
can easily be simplified to yield
the practically relevant expression for thin wafers with
homogeneous generation, i.e. $\alpha^{-1} \gg d,l$,
\begin{equation}
\tau^{\alpha \to 0}_{\rm eff} = 
\tau_{\rm b} \left[ 1 - \frac{2 \frac{S l}{D} \frac{l}{d}}{1+\frac{S l}{D} \coth\frac{d}{2l}} \right] \, .
\end{equation}
Furthermore, the case of very large surface recombination is given
by
\begin{equation}
\tau^{S \to \infty}_{\rm eff} = 
 \frac{\tau_{\rm b}}{1-\alpha^2 l^2} \left[ 1 -
 \alpha l \frac{ \tanh \frac{d}{2 l}}{\tanh \frac{\alpha d}{2}} \right]
\end{equation}
which extends the result for homogeneous carrier generation
$\tau^{\alpha \to 0, S \to \infty}_{\rm eff} = 
 \tau_{\rm b} \{ 1 - 2 (l/d) \tanh [d/(2 l)]\}$
presented in Ref.~\onlinecite{Sinton04b}.
For thick blocks with $d \gg l, \alpha^{-1}$ one finds
\begin{equation}
\tau^{d\to\infty}_{\rm eff} = 
 \frac{\tau_{\rm b}}{1-\alpha^2 l^2} \left[ 1- \alpha l
 \frac{\alpha l + S l /D}{1+S l / D} \right]
\end{equation}
which is a generalization of the result given in
Ref.~\onlinecite{Bowden07} as it includes the surface
recombination velocity $S$ explicitly.
It turns out that the restriction $S_0=S_1$ is not necessary to obtain
this result as only the front side of a block influences the total number
of excess carriers
significantly. Assuming the front side to be unpassivated, i.e. $S=S_0 \to \infty$,
the expression simply reproduces the result
$\tau^{d\to\infty, S\to\infty}_{\rm eff} \approx \tau_{\rm b}/(1+\alpha l)$ 
given in Ref.~\onlinecite{Bowden07}. Note, that the corresponding
surface recombination time $\tau_{\rm s}=l/(\alpha D)$
is not covered by the phenomenological expression $\tau_{\rm s,tr}$
derived in the transient case.

\begin{figure}
\parbox[t]{0.9\linewidth}
 {\includegraphics[width=\linewidth]{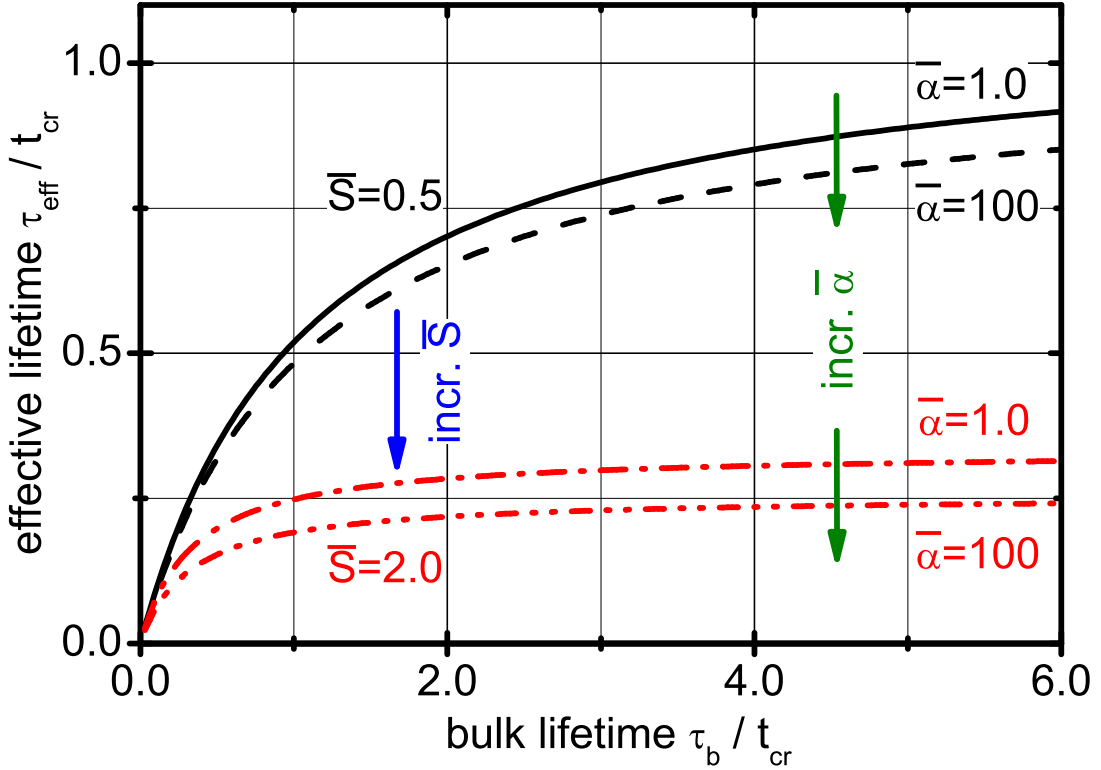}
  \caption{\label{fig:eff_lifetime_vs_bulk_S_alpha}(Color online) Measurable
  (effective) lifetime
  $\tau_{\rm eff}$ (in units of $t_{\rm cr}=d^2/D$) as a function of
  bulk lifetime $\tau_{\rm b}$ and for the dimensionless parameters
  $\bar S = 0.5$, $\bar \alpha = 1.0$ (solid, black),
  $\bar S = 0.5$, $\bar \alpha = 100$ (dashed, black),
  $\bar S = 2.0$, $\bar \alpha = 1.0$ (dash-dot, red),
  $\bar S = 2.0$, $\bar \alpha = 100$ (dash-dot-dot, red).
  }
 }
\end{figure}

Before the correct quantitative asymptotic limits are discussed in
more detail, the qualitative
features of relationship (\ref{eq:tau_eff_equal_S}) are studied
in terms of the dimensionless
parameters $\bar S \equiv Sd/D$, $\bar \alpha \equiv \alpha d$, and
$\bar l^2 = l^2/d^2 = \tau_{\rm b}/t_{\rm cr} \equiv \bar \tau_{\rm b}$.
Here, $t_{\rm cr} \equiv d^2/D$ stands for the time scale
defined by the diffusion through the entire sample. The dependence
of $\tau_{\rm eff}$ on $\tau_{\rm b}$ for various parameters $\bar S$ and
$\bar \alpha$ is shown in
Fig.~\ref{fig:eff_lifetime_vs_bulk_S_alpha}.
One can identify a region of small bulk lifetimes where
$\tau_{\rm eff} \approx \tau_{\rm b}$.
In this region, a measurement of $\tau_{\rm eff}$ allows a direct determination
of $\tau_{\rm b}$. The size of this region increases for lower surface recombination,
i.e. decreasing $\bar S$.
On the other hand, for large bulk lifetimes the value of $\tau_{\rm eff}$ saturates
even if $\tau_{\rm b}$ is increasing further. This implies that a measurement
of $\tau_{\rm eff}$ is not sensitive to any changes in the bulk property $\tau_{\rm b}$
anymore since the surface recombination dominates. Therefore, a measurement of
$\tau_{\rm eff}$ in this region can in principle not give any information about $\tau_{\rm b}$.
Finally, there is a region of intermediate values for $\tau_{\rm b}$
where a direct proportionality between $\tau_{\rm eff}$ and $\tau_{\rm b}$ is not given.
However, the information gained in a measurement of $\tau_{\rm eff}$ could still be used
to determine $\tau_{\rm b}$. In the next section,
these features will be discussed in a more formal and quantitative way.

\begin{figure}
\parbox[t]{0.9\linewidth}
 {\includegraphics[width=\linewidth]{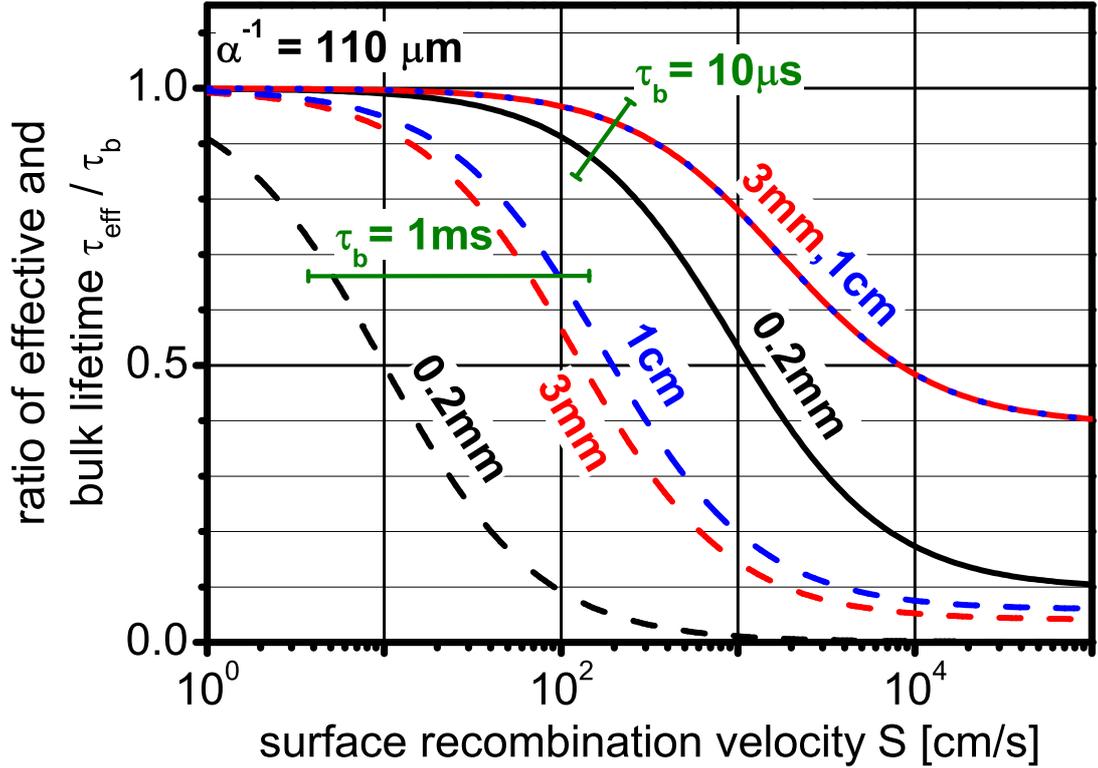}
  \caption{\label{fig:eff_lifetime_vs_S}(Color online) Ratio between
  measurable lifetime
  $\tau_{\rm eff}$ and
  bulk lifetime $\tau_{\rm b}$ as a function of the surface recombination
  velocity $S$. Shown are two different material qualitites: $\tau_{\rm b}=10\,{\rm \mu s}$
  (solid lines) and $\tau_{\rm b}=1\,{\rm ms}$
  (dashed lines). The thickness of the samples are indicated and
  increase from $d=200\,{\rm \mu m}$ (left curves, black)
  to $d=1\,{\rm cm}$ (right curves, blue).
  }
 }
\end{figure}

The influence of the surface passivation on the measurement is
presented in more detail in Fig.~\ref{fig:eff_lifetime_vs_S}.
As already noted, the measured lifetime is approximately equal to the
bulk lifetime when the surface recombination is sufficiently suppressed.
The importance of the surface passivation decreases with increasing
sample thickness and bulk recombination. Furthermore, there is another
important aspect to be seen in Fig.~\ref{fig:eff_lifetime_vs_S}. For
parameter regimes where $S$ is either small or large, the ratio
$\tau_{\rm eff} / \tau_{\rm b}$ becomes constant and thus significantly
less sensitive to the precise value of $S$. It implies
that the
bulk value $\tau_{\rm b}$ can be concluded from $\tau_{\rm eff}$
for large sample thickness and large
surface recombination even if the precise value of $S$
is not known. This justifies the approach in Ref.~\onlinecite{Bowden07}.
In the intermediate region (in Fig.~\ref{fig:eff_lifetime_vs_S}
roughly between $S\sim 10\,{\rm cm/s}$ and $S\sim 10^4 {\rm cm/s}$) a detailed
knowledge of the surface recombination velocity is necessary in order
to determine $\tau_{\rm b}$ from a measurement.

Compared to the strong dependence of the ratio $\tau_{\rm eff}/\tau_{\rm b}$ on
$\bar S$, the influence of $\bar \alpha$ is rather weak, see
Fig.~\ref{fig:eff_lifetime_vs_bulk_S_alpha}.
Nevertheless, this dependence can be employed by measuring
the same sample two or more times injecting carriers
with light of different wave lengths. These wavelength
dependent measurements allow elliminating
other unknown quantities such as $S$.\cite{Bail00,Brody02}

\begin{figure}
\parbox[t]{0.9\linewidth}
 {\includegraphics[width=\linewidth]{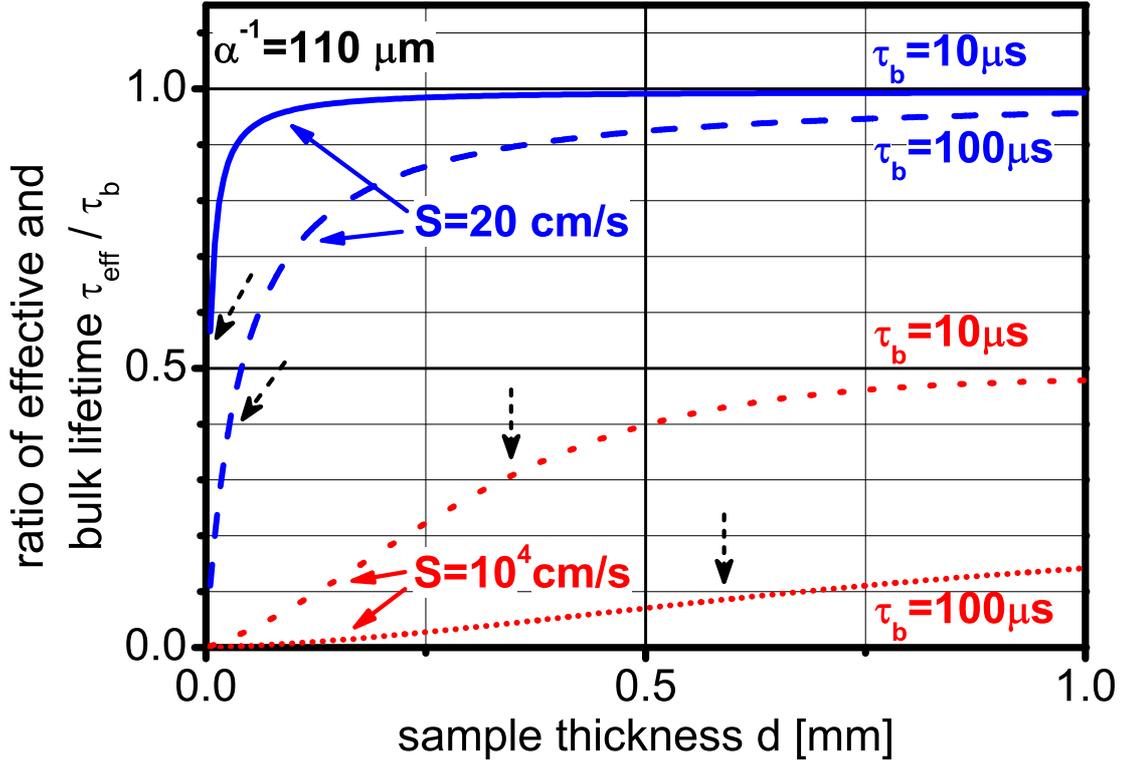}
  \caption{\label{fig:eff_lifetime_vs_thickness}(Color online) Ratio between
  measurable lifetime
  $\tau_{\rm eff}$ and
  bulk lifetime $\tau_{\rm b}$ as a function of the sample thickness $d$.
  Shown are two different material qualitites: $\tau_{\rm b}=10\,{\rm \mu s}$
  (solid and dotted line) and $\tau_{\rm b}=100\,{\rm \mu s}$
  (dashed and short dashed line). The surface recombination velocity is
  chosen to be $S=20\,{\rm cm/s}$ (upper curves, blue) and $S=10^4\,{\rm cm/s}$
  (lower curves, red). The absorption length is $\alpha^{-1}=110\,{\rm \mu m}$
  for all four cases. The dashed arrows indicate the critical thickness
  $d_{\rm crit}$, see Sec.~\ref{sec:asymptotic}.
  }
 }
\end{figure}

The sample thickness is the third relevant parameter.
In Fig.~\ref{fig:eff_lifetime_vs_S}, one can observe that 
for increasing thickness a limiting curve
is approached which means that an increase in $d$
beyond a certain critical thickness does not change the measurement result
much. This can also be seen from $\tau_{\rm eff}^{d\to\infty}$
which is independent of $d$.
The critical thickness is lower for materials with lower bulk lifetime.
The relation between $\tau_{\rm eff}$ and $\tau_{\rm b}$ as a function
of $d$ itself is presented in 
Fig.~\ref{fig:eff_lifetime_vs_thickness}. The ratio
$\tau_{\rm eff}/\tau_{\rm b}$ reaches a constant value smaller than one
for very large thickness, i.e. $d \to \infty$. This implies that although 
$\tau_{\rm eff}$ is always smaller than $\tau_{\rm b}$ there
is nevertheless a proportionality between these two quantities.
Therefore, it is possible to deduce the bulk property from a 
measurement of $\tau_{\rm eff}$ for samples of sufficient thickness.
This is different for the
limit of very small thickness, i.e. $d\to 0$. In this limit,
the measurable lifetime vanishes even for a finite value
of $\tau_{\rm b}$. This is caused by the strongly increased
impact of the surface recombination on the overall measurement.
Between these two limits there is a regime of intermediate
thickness where a change in $\tau_{\rm b}$ is reflected
in some change of $\tau_{\rm eff}$. The onset of this regime
can be estimated by a critical thickness $d_{\rm crit}$
beyond which a measurement of the effective lifetime allows for
a determination of the bulk lifetime. An analytic expression
for this critical thickness in terms of the surface recombination
$S$, the diffusion length $l$ (bulk lifetime $\tau_{\rm b}$)
and absorption length $\alpha^{-1}$
is derived in the Sec.~\ref{sec:asymptotic}. Qualitatively,
it is clear that a larger surface recombination velocity
and a larger bulk lifetime
lead to a larger critical thickness, see Fig.~\ref{fig:eff_lifetime_vs_thickness}.

\section{\label{sec:asymptotic}Asymptotic expressions}
Quantitative conclusions
can be drawn by considering various limits of 
Eq.~(\ref{eq:tau_eff_equal_S}) with respect to the dimensionless
surface recombination velocity $\bar S$,
the absorption length $\bar \alpha^{-1}$, and the
bulk lifetime $\bar \tau_{\rm b} = \bar l^2$.
In terms of these parameters, Eq.~(\ref{eq:tau_eff_equal_S}) can be rewritten as
\begin{equation}
\label{eq:tau_eff_F1_F2}
\tau_{\rm eff} = \tau_{\rm b} \left[ 1 - {\cal F}_1(\bar S, \bar l) \cdot 
 {\cal F}_2(\bar \alpha, \bar l) \right]
\end{equation}
with
\begin{equation}
{\cal F}_1 \equiv \frac{2 \bar S \bar l^2}{1+\bar S \bar l \coth \frac{1}{2\bar l}}
\,
\text{,}
\quad
{\cal F}_2 \equiv
\frac{\bar \alpha}{2} \, \frac{\coth \frac{\bar \alpha}{2} - \bar \alpha \bar l \coth \frac{1}{2\bar l}}
 {1-\bar \alpha^2 \bar l^2} \, 
\end{equation}
where ${\cal F}_1$ depends on $\bar S$ and $\bar l$ only while
${\cal F}_2$ is a function of $\bar \alpha$ and $\bar l$ only.
Using the result (\ref{eq:tau_eff_F1_F2}) and the
definition $\tau_{\rm s} = (\tau_{\rm eff}^{-1} - \tau_{\rm b}^{-1})^{-1}$
one finds
\begin{equation}
\label{eq:tau_s_F1_F2}
\tau_{\rm s} = \tau_{\rm s}(\bar l, \bar S, \bar \alpha)
 = \tau_{\rm b} \left[ {\cal F}_1^{-1}(\bar S, \bar l) \, {\cal F}^{-1}_2(\bar \alpha, \bar l) - 1 \right]
  \, .
\end{equation}
The measurement $\tau_{\rm eff}$ is hence dominated by the bulk property
$\tau_{\rm b}$ if $\tau_{\rm s}/\tau_{\rm b} \gg 1$, i.e. for
${\cal F}_1 {\cal F}_2 \to 0$. The other limit of
${\cal F}_1 {\cal F}_2 \to 1$ implies $\tau_{\rm s}/\tau_{\rm b} \ll 1$
which means that the surface properties dominate and
$\tau_{\rm eff} \approx \tau_{\rm s}$.

The Eqs.~(\ref{eq:tau_eff_F1_F2}) and (\ref{eq:tau_s_F1_F2}) can be
simplified for various applications by means of the approximation
\begin{equation}
x \coth\frac{x}{2} \approx
 \left\{ \begin{array}{l c l} x & {\rm for} & x \gg 3\\
  2 + x^2/6 - x^4/360 & {\rm for} & x \ll 3\end{array}
 \right. 
\end{equation}
together with standard series expansions.
In the intermediate regime between the two limiting cases, i.e. at $x\approx 3$,
this aproximation gives an error that is at most around 10\%. The
application of the approximation to ${\cal F}_1$ and ${\cal F}_2$ leads,
for example, in a direct way to the approximations of Eq.~(\ref{eq:tau_eff_F1_F2})
that are applicable for small and large sample thickness $d$, i.e.
\begin{eqnarray*}
{\cal F}_1(\bar S, \bar l) \approx \left\{ \begin{array}{l c l}
 2 \bar S \bar l^2/\left(1+\bar S [2 \bar l^2 + 1/6]\right) & \text{for} & \bar l > 0.3 \\
 2 \bar S \bar l^2 / \left(1+\bar S \bar l\right) & \text{for} & \bar l < 0.3
 \end{array} \right.
\end{eqnarray*}
and
\begin{eqnarray*}
{\cal F}_2(\bar \alpha, \bar l) \approx \left\{ \begin{array}{l c l}
 1 + \bar \alpha^2/\left(720 \bar l^2\right) & \text{for} &  \bar \alpha < 3, \bar l > 0.3 \\
 \bar \alpha/\left(2+2\bar \alpha \bar l\right) & \text{for} &  \bar \alpha > 3, \bar l < 0.3
 \end{array} \right. \, ,
\end{eqnarray*}
respectively.

To illustrate the approach based on
the two functions ${\cal F}_{1,2}$ and the applicability of the
result Eq.~(\ref{eq:tau_eff_F1_F2}), the following
practical examples for p-type Silicon are considered
(see Table~\ref{tab:cases_param}
for details on the chosen parameters):

{\em 1. Surface-passivated wafer of low to intermediate material quality} implying
 intermediate $\bar l$, small $\bar S$ and $\bar \alpha$: In this limit
 one finds ${\cal F}_1 \approx 2 \bar S \bar l^2$ and
 ${\cal F}_2 \approx 1 + \bar \alpha^2/(720 \bar l^2)$. Thus,
 according to the chosen parameters (see Table~\ref{tab:cases_param}),
 ${\cal F}_1 \cdot {\cal F}_2 \ll 1$ is a small correction
 in Eq.~(\ref{eq:tau_eff_F1_F2}) which in this limit then reads
 $\tau_{\rm eff} \approx \tau_{\rm b}[1-2 \bar S \bar l^2]$.
 This relation describes the curves shown in Fig.~\ref{fig:eff_lifetime_vs_S}
 for small $S$. Furthermore one can deduce the surface recombination time
 which is given by
 $\tau_{\rm s} \approx \tau_{\rm b} / ({\cal F}_1 {\cal F}_2) \approx d/(2S)$.
 Therefore, one obtains the same result as for the transient approach
 if the leading order in the small correction is considered.

{\em 2. Wafer of low to intermediate material quality
 without surface passivation} characterized by
 intermediate $\bar l$, small $\bar \alpha$, and large $\bar S$: In this
 case the asymptotic expressions are
 ${\cal F}_1 \approx 2 \bar S \bar l^2/(1+2\bar S \bar l^2 + \bar S / 6)$ and
 ${\cal F}_2 \approx 1 + \bar \alpha^2/(720 \bar l^2)$. In contrast to
 case 1, $\bar S$ is not small implying that
 ${\cal F}_1 \cdot {\cal F}_2$ is not a small correction
 in Eq.~(\ref{eq:tau_eff_F1_F2}) and therefore
 the measured $\tau_{\rm eff}$ is much smaller than the bulk
 value $\tau_{\rm b}$. This is the behaviour
 presented in Fig.~\ref{fig:eff_lifetime_vs_S} for large $S$.
 Qualitatively, this is not
 surprising as the measured lifetime is governed by the large
 surface recombination. Practically, it allows for an estimation
 of an upper bound of the surface recombination velocity $S$
 by assuming $\tau_{\rm s} \approx \tau_{\rm eff}$. The surface recombination time
 $\tau_{\rm s}$ relates to $S$ via
 $\tau_{\rm s} \approx d/(2S) \cdot [1+Sd/(6D)]$
 which
 is found from Eq.~(\ref{eq:tau_s_F1_F2}). Since $S$ is not small, the second term
 in the brackets cannot be neglected as in the first case.
 The inverse relation
 $S\approx d/(2\tau_{\rm s}) \, [1-d^2/(12 D \tau_{\rm s})]^{-1}$
 can then be employed to calculate $S$.

{\em 3. Surface-passivated float-zone wafer (very high material quality)} with
 large $\bar l$, small $\bar S$, and small $\bar \alpha$: The asymptotic
 behaviour of ${\cal F}_1$ and ${\cal F}_2$ is analogous to
 the previous case, i.e.
 ${\cal F}_1 \approx 2 \bar S \bar l^2/(1+2\bar S \bar l^2 + \bar S / 6)$ and
 ${\cal F}_2 \approx 1 + \bar \alpha^2/(720 \bar l^2)$,
 leading to the same expressions for $\tau_{\rm eff}$ and $\tau_{\rm s}$
 when inserted in Eqs.~(\ref{eq:tau_eff_F1_F2}) and (\ref{eq:tau_s_F1_F2}).
 Again, the measurement of $\tau_{\rm eff}$ is dominated by
 surface effects allowing for the estimation of an upper bound
 of the surface recombination velocity $S$.

{\em 4. Block of low to intermediate material quality
 without surface passivation} describable by
 small $\bar l$, large $\bar S$ and $\bar \alpha$: The
 asymptotics in this case read
 ${\cal F}_1 \approx 2 \bar S \bar l^2/(1+\bar S \bar l)$ and
 ${\cal F}_2 \approx \bar \alpha/(2 + 2 \bar \alpha \bar l)$.
 Together with Eq.~(\ref{eq:tau_eff_F1_F2}) this exactly reproduces the result
 $\tau_{\rm eff}^{d\to\infty}$.
 Again, the product ${\cal F}_1 \cdot {\cal F}_2$ is not a small
 correction. Nevertheless, the bulk information can be extracted
 from the measurement.
 
\begin{table}[!ht]
\begin{tabular}{|c||c|c|c||c|c|c||c|c|}
\hline
case &  $\tau_{\rm b}$ & $d$ &  $S$
     &  $\bar l = \frac{l}{d}$ & $\bar S = \frac{Sd}{D}$ & $\bar \alpha = \alpha d$
     &  $\tau_{\rm eff}$ & $\tau_{\rm s}$ \\
     &  $[\un{\mu s}]$ & $[\un{mm}]$ & $[\un{cm/s}]$ & & & & $[\un{\mu s}]$ & $[\un{\mu s}]$ \\ \hline
1  & $10$   & $0.20$ &  $20$   & $0.9$  & $0.01$ & $1.8$ & 9.8 & 490 \\ 
2  & $10$   & $0.20$ &  $10^4$ & $0.9$  & $6.7$  & $1.8$ & 1.7 & 2.1 \\ 
3  & $1000$ & $0.20$ &  $20$   & $8.7$  & $0.01$ & $1.8$ & 334 & 502 \\ 
4  & $10$   & $10$   &  $10^4$ & $0.02$ & $333$  & $91$  & 4.8 & 9.2 \\
\hline
\end{tabular}
\caption{\label{tab:cases_param}
 Dimensionless parameters $\bar l$, $\bar S$, and $\bar \alpha$
 for the examples under consideration. The absorption length
 has been chosen to be constant for all cases, i.e.
 $\alpha^{-1} = 110\,{\rm \mu m}$), while the diffusion constant
 is set to $D=30\,{\rm cm^2/s}$.}
\end{table}

It is clear from the previous discussion and 
from Fig.~\ref{fig:eff_lifetime_vs_bulk_S_alpha} that a measurement
of a sample of lower material quality is more likely to give
an accurate estimate of the bulk property $\tau_{\rm b}$. This is reflected
by the fact that $\tau_{\rm eff}/\tau_{\rm b} \to 1$ for $\tau_{\rm b} \to 0$.
However, in most pratical cases one cannot expect the
relation $\tau_{\rm eff} = \tau_{\rm b}$ to be exact and 
the corrections to this equality are of interest. Using the
asymptotics of ${\cal F}_1\approx 2 \bar S \bar l^2$ and
${\cal F}_2\approx \bar \alpha [\coth(\bar \alpha/2) - \bar \alpha \bar l]/2$
yields
\begin{eqnarray*}
\tau_{\rm eff}^{\tau_{\rm b} \to 0} & = &
\tau_{\rm b} \left[ 1 - \bar l ^2 \, 
 \bar S \bar \alpha \coth \frac{\bar \alpha}{2} \right] \\
 & \approx & \left\{ \begin{array}{lcl}
 \tau_{\rm b} \left[ 1 - 2 \bar S \bar l^2 \right] & {\rm for} & \bar \alpha \ll 1\\
 \tau_{\rm b} \left[ 1- \bar \alpha \bar S \bar l^2 \right] & {\rm for} & \bar \alpha \gg 1
\end{array} \right.
\end{eqnarray*}
indicating the first order correction in $\bar \tau_{\rm b} \equiv \bar l^2$
being valid for material of poor quality (low bulk lifetimes).
The criterion for how large the bulk lifetime of a sample
of fixed thickness and surface quality can become before
a measurement of $\tau_{\rm eff}$ does not relate to bulk properties anymore
is hence given by the correction being much smaller than one.
There is a rather intuitive interpretation for this. For
homogeneous generation, i.e. $\bar \alpha \ll 1$,
the condition $2\bar S \bar l^2 = 2 S \tau_{\rm b}/d = \tau_{\rm b}/\tau_{\rm s} \ll 1$
just means that the two regions where the surface recombination
dominates are of thickness $2S \tau_{\rm b}$ and have to be negligible
compared to the total sample thickness $d$.
For strongly inhomogeneous generation, i.e. $\bar \alpha \gg 1$,
the criterion is replaced by
$\bar \alpha \bar S \bar l^2 = S \tau_{\rm b} \alpha \ll 1$
meaning that the width $S \tau_{\rm b}$ of the one region with dominating
surface recombination has to be smaller than the absorption
depth $\alpha^{-1}$.

On the other hand, the measured lifetime of a sample
of very high material quality, i.e. very large
bulk lifetime, is mostly dominated by the surface effects. In particular,
this means that $\tau_{\rm eff}^{\tau_{\rm b}\to \infty}$ saturates
at a value that is independent of $\tau_{\rm b}$, i.e.
$\tau_{\rm eff}^{\tau_{\rm b} \to \infty} \to \tau_{\rm s}$.
With the appropriate asymptotic expressions for
${\cal F}_1$ and ${\cal F}_2$ one finds
\begin{equation}
\label{eq:tau_eff_large_tau_b}
\tau_{\rm eff}^{\tau_{\rm b} \to \infty}
= \tau_{\rm s}^{\tau_{\rm b} \to \infty}
= t_{\rm cr} \frac{\bar \alpha^2 - 2 \bar S + \bar S \bar \alpha
\coth \frac{\bar \alpha}{2}}{2 \bar S \bar \alpha^2}
\end{equation}
In other words, a measurement on a sample of very high
material quality gives information on the quality of the
surface passivation, i.e. on $\tau_{\rm s}$,
rather than on $\tau_{\rm b}$. The limit
(\ref{eq:tau_eff_large_tau_b}) can thus be used to
determine $S$ for a given thickness $d$ and absorption length $\alpha^{-1}$.
This limit together with the general solution (\ref{eq:tau_eff_F1_F2})
suggests the large $\bar l \gg 1$ structure
$\tau_{\rm eff}^{\bar l \gg 1} = \tau_{\rm b}
[1- A^* \, 2 \bar l \tanh (1/\{2 \bar l\})]$
with $A^*=1-\tau_{\rm eff}^{\tau_{\rm b} \to \infty}/(t_{\rm cr} \bar l^2)$
which is slightly different from the
phenomenological proposal for unpassivated wafers
in Ref.~\onlinecite{Sinton04b} with respect to the
prefactor $A^*$.
In the special case of thin samples with homogeneous
generation, i.e. wafers with
$\bar \alpha \to 0$, the limit
(\ref{eq:tau_eff_large_tau_b}) reduces to
\begin{equation}
\tau_{\rm eff}^{\tau_{\rm b} \to \infty, \alpha \to 0}
= \tau_{\rm s}^{\tau_{\rm b} \to \infty, \alpha \to 0} =
t_{\rm cr} \frac{6+\bar S}{12 \bar S} = \frac{d}{2 S} + \frac{d^2}{12 D} \,.
\end{equation}
Note, that this result for the steady state approach
is similar but not equal to the surface recombination approximation
$\tau_{\rm s, tr}$ obtained in the transient case. The difference
lies in the numerical prefactor of the term $\sim d^2$ that
is relevant for samples of larger thickness.

Experimentally, an adjustment of the bulk lifetime is typically more
difficult than the preparation of a sample with a specified thickness.
Therefore, the approximate behaviour in the limit of small
and large sample thickness is of particular interest.
The ratio $\tau_{\rm eff}/\tau_{\rm b}$ saturates at a constant
value smaller than one for very thick samples, see
Fig.~\ref{fig:eff_lifetime_vs_thickness}.
In particular, one finds with the asymptotics
${\cal F}_1 \approx 2 \bar S \bar l^2/(1+\bar S \bar l)$ and
${\cal F}_2 \approx \bar \alpha / (2 + 2\bar \alpha \bar l)$
the result
\begin{equation}
\label{eq:tau_eff_large_d}
\frac{\tau_{\rm eff}^{d\to\infty}}{\tau_{\rm b}} \approx
 1 - \frac{\bar S \, \bar \alpha \, \bar l^2}
 {(1+\bar S \bar l)(1+\bar \alpha \bar l)} =
1 - \frac{\alpha l}{1+\alpha l} \frac{S \, l}{D + S l}
\end{equation}
which is exactly the result presented in Sec.~\ref{sec:application}
for blocks. This relationship allows for a 
determination of $\tau_{\rm b}$ as a function of
$\tau_{\rm eff}$. In general, the knowledge
of $S$ would be required in order to apply this relation.
However, it becomes clear from Eq.~(\ref{eq:tau_eff_large_d})
that for unpassivated blocks with large $S$ the influence of
this parameter is rather weak and can thus be neglected.
On the other hand, the
ratio $\tau_{\rm eff}/\tau_{\rm b}$ vanishes
for small thicknesses which is caused by the 
surface effects. The asymptotic behaviour
for $d\to 0$ is obtained from $\bar \alpha, \bar S \ll 1$
and $\bar l \gg 1$ giving
${\cal F}_1 \approx 2 \bar S \bar l^2/[1+\bar S(2\bar l^2 + 1/6)]$
and
${\cal F}_2 \approx 1 + \bar \alpha^2/(720 \bar l^2)$.
Quantitatively, one thus finds
\begin{equation}
\label{eq:tau_eff_small_d}
\frac{\tau_{\rm eff}^{d\to 0}(d)}{\tau_{\rm b}} \approx
1 - \frac{2 \, \bar S \, \bar l^2}{1+2\bar S \bar l^2 + \bar S/6}
= 1-\frac{2 \, S \, l^2}{D d + 2 S l^2 + S d^2/6}
\end{equation}
which is correct up to the second order in $d$.
Hence, a determination of $\tau_{\rm b}$ from
$\tau_{\rm eff}$ is not possible in this
regime since $\tau_{\rm eff} \approx d/(2S)$
for $d \to 0$ depends only on $S$ but not on $\tau_{\rm b}$.
Therefore, the lifetime measurement on thin wafers can be
used to determine $S$ but it does not allow
to deduce the bulk property $\tau_{\rm b}$.
The critical thickness that separates the two regimes of
large thickness, Eq.~(\ref{eq:tau_eff_large_d}),
and small thickness, Eq.~(\ref{eq:tau_eff_small_d})
can be estimated by 
$\tau_{\rm eff}^{d \to \infty} / 2 = \tau_{\rm eff}^{d \to 0}(d_{\rm crit})$,
see Fig.~\ref{fig:eff_lifetime_vs_thickness}.
Using Eq.~(\ref{eq:tau_eff_large_d}) and Eq.~(\ref{eq:tau_eff_small_d})
the corresponding quadratic equation in $d_{\rm crit}$
reads
\begin{equation}
d_{\rm crit}^2 + \frac{6 D}{S} d_{\rm crit} + 12 l^2 \frac{\alpha l^2 S - (1+\alpha l)(D + S l)}
 {(1+\alpha l)(D + S l) + \alpha l^2 S} = 0
\end{equation}
which can easily be solved. The two most interesting
cases are given by small $S$ (passivated samples) and
large $S$ (unpassivated samples) for which
the solutions are
\begin{equation}
\label{eq:d_crit}
d_{\rm crit}^{S\to 0} \approx 2 \, \tau_{\rm b} \, S
\quad \text{and} \quad
d_{\rm crit}^{S\to \infty} \approx \sqrt{6 \frac{l}{\alpha}} \, ,
\end{equation}
respectively. There is an intuitive interpretation of these
results. For samples with a good surface passivation
a measurement of the bulk properties is possible
if the sample is more than twice the width $\tau_{\rm b} S$
of the regions where the surface recombination takes place.
This picture cannot be applied in the case of low surface
passivation with large $S$ since the near-surface regions of width
$\tau_{\rm b} S$ would extend throughout the entire sample.
In this case, other length scales are important, in particular
the diffusion length $l$ and the absorption length $\alpha^{-1}$.
The critical thickness is then simply the geometric mean
of these two length scales.

\section{\label{sec:conclusions}Conclusions}
In this work, the generalized solution for the relation
between the measured lifetime and the bulk lifetime
of excess minority carriers is investigated. This relation
allows to study the interplay of the bulk and
surface recombination for a steady-state measurement.
It provides a general theoretical background
for a consistent data analysis of lifetime measurements
on a wide variety of samples.
In particular, it allows identifying the influence of the
relevant parameters, i.e. bulk lifetime $\tau_{\rm b}$,
surface recombination velocity $S$, absorption depth $\alpha^{-1}$,
and sample thickness $d$ on the measured lifetime
$\tau_{\rm eff}$. The practically most important
cases of passivated and unpassivated wafers and blocks
are discussed within the same framework 
and exact analytic relations for these
cases are presented. The proposed approach is
demonstrated for typical and practical relevant examples.
It is shown that a reliable measurement
of the bulk properties for given surface conditions is possible
only if the sample is thick enough and the surface passivation sufficiently
effective. On the other hand, lifetime
measurements on samples thinner than a critical thickness $d_{\rm crit}$
are best employed when the quality of the surface passivation layer
is of interest. For an unambiguous determination
of both $\tau_{\rm b}$ and $S$ at least two samples of intermediate
thickness have to be measured. Alternatively,
an experimental setup where only one sample is analyzed
with a light source of two different wave lengths, i.e.
two different $\alpha$, can give this information as well.
These qualitative conclusions are substantiated
by the corresponding quantitative relations
which allow an accurate analysis of experimental data.
Among those, an estimate of the relevant
critical thickness in dependence on the passivation, i.e. value
of $S$, and illumination, i.e. value of $\alpha^{-1}$, is derived.

\begin{acknowledgments}
The author acknowledges the funding of
the "Ministerium f{\"u}r Wirtschaft und Arbeit des Landes Sachsen-Anhalt"
within the ESCA Project (Project FuE 62/09).

\end{acknowledgments}
%

\end{document}